# AN INTEGRATED DESIGN OF ENERGY AND INDOOR ENVIRONMENTAL QUALITY MONITORING SYSTEM FOR EFFECTIVE BUILDING PERFORMANCE MANAGEMENT

*Vincent Gbouna Zakka* [1, 2] *& Minhyun Lee* [1]

[1] *Department of Building and Real Estate, The Hong Kong Polytechnic University*

[2] *College of Engineering and Physical Sciences, Aston University.*

***ABSTRACT:*** *Understanding the energy consumption pattern in the built environment is invaluable for the evaluation of the sources of energy wastage and the development of strategies for efficient energy management. An integrated monitoring system that can provide high granularity energy consumption and indoor environmental quality (IEQ) data is essential to enable intelligent, customized, and user-friendly energy management systems for end users and help improve building system performance. This paper, therefore, presents an integrated design of an internet of things (IoT)-based embedded, non-invasive, and user-friendly monitoring system for efficient building energy and IEQ management. The hardware unit of the system is comprised of a wireless microcontroller unit, current and voltage sensor, IEQ sensors, and power management unit that enables the acquisition, processing, and telemetering of energy and IEQ data. The software unit is made up of embedded software and a web-based information provision unit that handles real-time data analysis, transmission, and visualization on the web and provides relevant notifications to the end users about energy consumption patterns. The proposed system provides a promising solution for real-time information exchange regarding energy consumption and IEQ for both end-users and managers that will enhance effective building energy and environmental management.*

***KEYWORDS:*** *Building Energy Monitoring; Indoor Environmental Quality (IEQ) Monitoring; Internet of Things (IoT); Non-invasive System; Building Energy Management.*

## 1. INTRODUCTION

Buildings are responsible for over 40% of the global energy use and contribute towards 30% of the total $CO_2$ emissions (Ahmad, Mourshed, Mundow, Sisinni, & Rezgui, 2016). They consume the largest energy in the world and account for over one-third of total final energy consumption (Li et al., 2020). The fact that people spend more than 80% of their time in buildings has a significant impact on the usage of electrical appliances, space conditioning, and other building services, which also largely contribute to building energy consumption (Wagner A., 2018). Due to the amount of time people spend in buildings, the daily global $CO_2$ emissions from fossil fuels and consequently total energy consumption has increased (Le Quéré et al., 2020). As a result, improving building energy efficiency and reducing building energy consumption are becoming an increasingly important research area.

Cultural, societal, and social contexts in different parts of the world and the technologies that provide thermal comfort contribute to the difference in levels of energy consumption across buildings, but also to a significant extent, the occupant behavior in buildings (S. Yang et al., 2020). The energy consumption in the built environment is directly connected to the occupant behavior in the building. Understanding the fine-grained occupancy information in the building environment is, therefore, an important parameter for efficient use of energy. The significant discrepancy between building energy use as designed and the actual operation shows a need to evaluate the relationship between building occupants and energy requirements (Naylor, Gillott, & Lau, 2018). With the current trend of designing and operating more energy-efficient buildings, one of the identified leading influences on building energy consumption is occupant behavior. The tight requirements towards building energy performance and sustainability have given rise to awareness of the importance of understanding building occupants' presence and behavioral patterns (Esfandiari et al., 2021).

Unaware behaviors of occupants have great impact on energy consumption in the built environment. Real-time data and timely information of the actual energy consumption are required to reach an adequate awareness of energy consumption. Real-time feedback can be a powerful impetus for behavioral change as shown by previous studies that examine the effect of energy feedback information on occupant behavior. This was first reported in (McClelland & Cook, 1979) where the impact of continuous energy feedback information on energy usage was tested. Electricity usage was found to be lowered by 12% in homes with continuous electricity usage feedback compared to the homes with no usage of a feedback information system (Allen & Janda, 2006). To therefore enhance the energy efficiency of buildings, an information system aimed at increasing the awareness and engagement of end-users through frequent and detailed information on energy consumption is required (Sælen & Westskog, 2013). Lack of information to the end-users on energy consumption could lead to high energy consumption discrepancy between two buildings with similar thermos-physical characteristics and energy performance (Sonderegger, 1978). Collecting data on and evaluation of energy performance are at the core of the various strategies to enhance energy efficiency and to reduce energy use in buildings. A comprehensive energy metering monitoring system that will provide holistic understanding of energy consumption pattern has the potential to bring on board occupants and building owners to take energy-efficiency measures towards energy conservation (Genet & Schubert, 2011). Therefore, energy monitoring system that will provide information not only to engineers and utility companies but also to end users to optimize energy consumption is paramount.

In this regard, potential solutions for building energy monitoring have been proposed. In Suryadevara et al. (Suryadevara, Mukhopadhyay, Kelly, & Gill, 2015), the authors reported a smart and controlling system for household electrical appliances. The system was designed principally to monitor the electrical parameters of household appliances. Park et al. (Park, Choi, Kang, & Park, 2013) proposed a smart energy management system that functions as a control using a motion sensor and setting time of power usage to reduce power consumption. An embedded in-house energy information system with a smart energy controller was proposed in Kunold et al. (Kunold, Kuller, Bauer, & Karaoglan, 2011) which allows displaying of real-time data analysis of power consumption as well as power generation. Abdel-Majeed et al. (Abdel-Majeed, Tenbohlen, & Ellerbrock, 2013) presented the design of a cheap meter that could measure consumers' power consumption and generation along with control functions. A new commercial building energy management system using IoT based smart compact energy meter was proposed in Karthick et al. (Karthick, Raja, Nesamalar, & Chandrasekaran, 2021) to monitor and control energy quality issues. In Sánchez-Sutil et al. (Sánchez-Sutil, Cano-Ortega, & Hernández, 2021) a gateway for residential electricity metering networks using LoRa and an electrical variable measuring device for households was proposed to allow the development of smart meter networks with large coverage and low consumption. Sayed et al. (Sayed, Hussain, Gastli, & Benammar, 2019) presented the design of a modular and open-source smart meter for educational and research purposes. The aforementioned solutions have contributed to the design of building energy monitoring system. However, they suffer from some limitations thereby leaving room for more research effort to realize a simple but yet effective energy monitoring system that will satisfy consumer's need and conserve energy. Most of the previously developed smart meters are invasive which will require an electrician for installation since there will be need to do some high voltage electrical work. In addition, there were no clear description of system design for some of the proposed systems and are therefore not reproducible. A practical method to use the energy data to help end users to make timely and informative decision in conserving energy consumption is still lacking. More so, some of the proposed systems are complex and cumbersome and may not be desirable for most consumers since they do not have the requisite technical knowledge. A relatively simple system design was proposed in Z. Yang et al. (Z. Yang et al., 2018), however, the interfacing with electrical system for energy monitoring was not clearly explained thereby making it impracticable for a consumer without technical knowledge.

In addition, to not only promote energy-saving behavior but also maintain healthy indoor environment at the same time, real-time building energy monitoring supplemented by relevant indoor environmental quality (IEQ) information is significant. Monitoring energy consumption alone can be useful to achieve financial benefits, however, a holistic monitoring of the performance of the building system is required to identify the factors influencing irregular energy usage and for efficient building management. Relevant information regarding irregularity of building system performance will be essential to mitigate energy wastage and provide the required information needed to encourage building occupant behavioral change. Granular energy data and IEQ information would translate into increasing energy use awareness, and identification of factors influencing irregular energy usage in the built environment which can lead to a more sustainable building operation. Although several potential systems to monitor IEQ parameters have been proposed in the literature (Jin, Liu, Schiavon, & Spanos, 2018; Karami, McMorrow, & Wang, 2018; Mujan, Licina, Kljajić, Čulić, & Anđelković, 2021; Saini, Dutta, & Marques, 2020a, 2020b), they are however proposed as standalone systems designed to monitor only IEQ parameters. However, for a cost-effective, user-friendly, easy to deploy system that will provide a comprehensive understanding for formulation of strategies towards efficient energy conservation without compromising occupants comfort, an integrated solution comprising of energy and IEQ monitoring is imperative.

To address these challenges, this study aims to develop and propose a user-friendly, cost-effective, and portable energy and IEQ monitoring system, which can provide real-time feedback to occupants for enhancing occupant awareness. It is an integrated solution from hardware to software to allow efficient energy management through energy and IEQ monitoring in the built environment. Since understanding the patterns and habits of energy consumption are key requisites for energy conservation in buildings, the proposed system seeks to keep users updated about their energy consumption pattern and to offer them helpful tips for energy conservation. Monitoring of IEQ parameters namely air temperature, humidity, and air quality shall help to identify the relationship between energy consumption and occupant comfort. The system can be used to monitor both the whole building energy consumption and energy consumption by appliances when installed at the main electric board and at appliances level respectively. This will allow identification of higher consumption appliances, detection of abnormal conditions in current and voltage thereby helping end users to better manage energy consumption in the building. The integrated system has the benefit of unified platform for analyzing the patterns of energy consumption and IEQ which will serve as a useful tool for mitigating technical problems arising from real-time analysis of data from different platforms. The proposed system can be used to increase both the energy performance and occupant comfort in the built environment.

## 2. SYSTEM OVERVIEW

### 2.1 System architecture

The energy and IEQ monitoring system is an integrated device that can measure, analyze and communicate data to keep the end user updated about their energy consumption pattern and possible factors influencing the consumption rate. It is an infrastructure that includes energy monitoring system, IEQ monitoring system, wireless microcontroller unit (WMCU), power management system and information provision system as shown in Figure 1. It is a topology based on hybrid device that allows telemetering of energy consumption and IEQ data through the combination of measurements, processing, and recordings through communication networks. Each subsystem plays an indispensable role in the operation of the entire system. The WMCU is used for computation of the acquired data from energy and IEQ monitoring units. It also provides communication interface for data exchange with other systems which gives the flexibility of interfacing other devices for further expansion of the functionality of the proposed system to suit other application need. Furthermore, it provides the internet connectivity functionality that allows data communication over the network. The power management system ensures that the appropriate voltage and current for the proper functioning of the entire system is met from the source to the destination. The energy monitoring unit acquires the voltage and current signals through voltage and current sensors. The IEQ monitoring unit monitors the air temperature, humidity, and air quality through two environmental sensors. The acquired energy and IEQ signals are then conditioned, processed, and digitized through the analog to digital converter in the WMCU. The final data is then transmitted to the information provision system for storage, analysis, visualization and updating of the end user to create awareness about energy consumption pattern.

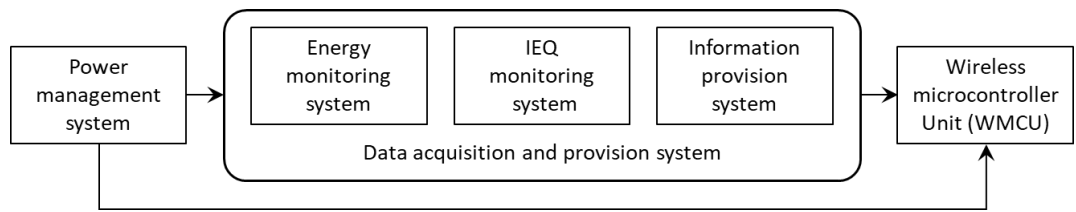

Fig. 1: System Architecture

### 2.2 System description

#### 2.2.1 Energy monitoring system

The energy monitoring unit is a communicating, non-invasive electrical meter that measures power consumption and communicate the data via Wi-Fi to the database. The proposed system is made up of a split-core current sensor for measuring current, AC-AC voltage adapter for measuring voltage, and signal conditioning circuit. The split-core current sensor is useful for measuring both whole building and appliance electricity consumption conveniently by clipping it onto either the live or neutral wire without the need to do any high voltage electrical work. This non-invasive method of measuring the current is particularly ideal for end users without deep technical knowledge. The alternating current (AC) voltage measurement by the AC-AC voltage adapter allows safe measurement of voltage which requires no high voltage work. The transformer in the adapter provides isolation from the high voltage mains. The signal conditioning circuit conditions the outputs from the current and voltage sensors to meet the input requirements of the WMCU. The energy monitoring unit is then interfaced with the WMCU for data processing, analysis and transmission. Based on the current and voltage readings, the WMCU then calculates active power, root mean square (RMS) voltage, and RMS current.

**2.2.2   IEQ monitoring system**

The IEQ monitoring unit comprise of air temperature and humidity sensor (AM2320), and air quality sensor (MQ135) for measuring environmental data to analyze its relationship with energy consumption. The AM2320 is a digital temperature and humidity sensor with a calibrated digital signal output. The sensor consists of a capacitive moisture element and an integrated high-performance temperature measurement device. Its fast response and anti-interference ability make it suitable for environmental monitoring. MQ135 is an air quality sensor suitable for detecting $CO_2$ concentration. The sensor's sensitive material is Tin(IV) oxide ($SnO_2$) which has low conductivity in clean air. When the target pollution gas exists, the sensor's conductivity gets higher along with the gas concentration rising. The sensor's capability to measure $CO_2$ concentration accurately makes it a good candidate for air quality monitoring.

**2.2.3   Power management system**

The power management system helps to ensure safe, reliable, and efficient operation of the whole system. It regulates and delivers the appropriate current and voltage for the proper functioning of each of the subsystems. It comprises of voltage regulators along with other electrical components for power regulation, removal of power distortion and noise, and protect the whole system by supplying pure direct current (DC). Any DC source can serve as an input source for the system. This means that the system can be powered either by battery or with an AC-DC voltage adapter from wall socket.

**2.2.4   Wireless microcontroller unit (WMCU)**

The WMCU is one of the core subsystems in the proposed system. It provides interface for communication between the various subsystems. Because of the requirement of wireless communication of the monitored data with the database, a microcontroller with wireless capability was chosen for this study. Also, to keep the overall size of the whole system compact, a microcontroller with a self-contained wireless networking solution along with input and output ports to provide interface for communication with other devices was desirable. Therefore, ESP32, a low-cost, dual core, low-power, system on chip (SoC) with Wi-Fi and dual-mode Bluetooth capabilities was chosen. The integration of Bluetooth, Bluetooth LE, and Wi-Fi capabilities gives room for a wide range of intuitive and user-friendly application development. The WMCU acquires the data from the energy and IEQ monitoring units to compute, analyze and transmit it to the database.

**2.2.5   Information provision system**

The information provision system provides real-time feedback to the end users about their energy consumption pattern and any irregularity about energy consumption rate. To keep the end users well informed with their energy use and indoor environment, instantaneous and cumulative energy and IEQ data is made available via a web-based personalized dashboard and an email system. The information is given to the end users for the purpose of minimizing energy consumption while maintaining their comfort. With the help of feedback and suggestions, the end user is made aware of the actual energy consumption and by involving the end users in the process, efficient energy use can be achieved. With the information provision system, the end user's energy-saving and healthy living can be improved.

## 3.   DESIGN PROCESS

## 3.1   Hardware design

### 3.1.1   Energy monitoring system design

**(1)   Current sensor design**

The current sensor used in this study for measuring AC current is a split core current transformer (CT) which allows non-invasive current measurement. It comprises of primary winding, magnetic core and a secondary winding. To measure the AC current, a wire carrying an electric current is passed through the opening of the sensor. This wire then forms the primary winding of the transducer. The secondary winding is made of several turns of wire enclosed within the transformer case. The AC current flowing in the primary winding then produces a magnetic field in the core which in turns induces a current in the secondary winding. To interface the transducer with WMCU, the output signal needs to be conditioned to meet the input requirement of the WMCU since the output from the sensor is higher than the input requirement. Firstly, the current signal from the sensor is converted to a voltage signal using a burden resistor. The burden resistor's value is chosen to provide a voltage that is proportional to the secondary current. The burden resistor was calculated using the formula in Eq. (1).

$$Burden\ Resistor = \frac{AREF \times CT\ turns}{2\sqrt{2} \times \max primary\ current} \tag{1}$$

Where $AREF$ is the Analog Reference voltage of the WMCU.

Next, a voltage divider bias circuit was designed to ensure that only the positive voltage signal is fed into the WMCU input. The capacitor in the bias circuit helps to eliminate noise which could be coupled into the circuit. The output signal is then interfaced with the analog/digital (A/D) converter input port of the WMCU to be converted into digital values as shown in Figure 2.

**(2) Voltage sensor design**

For a user-friendly measurement, AC-AC voltage adapter was used as the voltage sensor for voltage measurement. The output waveform of the sensor needs to be converted to a waveform that will meet the input the requirement of the WMCU. As shown in Figure 2, the output voltage signal was scaled down using a voltage divider circuit connected across the adapter's terminal. Also, a voltage divider bias circuit was designed to ensure that the input to the WMCU is a positive voltage signal. The scaling down of the waveform and the bias circuit ensures that the output signal from the adapter satisfies the input requirement of the WMCU. The conditioned output signal is then fed into the WMCU A/D converter input for conversion of the raw signal into digital values.

### 3.1.2 IEQ monitoring system design

For convenient interfacing of the IEQ monitoring system with the main system, breakout board (module) with the environmental sensors was used to provide easy plug and play experience. Interface was therefore provided in the form of pin headers on the main system board where the sensor modules with pins can be easily inserted. This option was preferred over integrating the sensors directly with the main system board because it is more user-friendly and gives the end users the flexibility of using other types of environmental sensors and not limiting them to a particular type. As shown in Figure 2, the core components for the air temperature and humidity sensor module are the AM2320 sensor, pin header, and pull-up resistors, while those for the air quality sensor module are the MQ135 sensor, potentiometer, a voltage comparator (LM393), and the pin header. By simply inserting the modules into the interface provided on the main board system, they can be detected by the WMCU and the output signal can be read out for further processing.

### 3.1.3 Power management system design

The power management system supplies the whole system with the required power supply. Its design is very crucial since the different subsystems have different power input requirement and each requirement should be satisfied for proper functioning. To therefore meet the input requirement of all the subsystems, the power management system was carefully designed to have multiple output power sources as shown in Figure 2. In addition, since one of the important requirements of the whole system is to be user-friendly, the power management system was designed to enable end users to easily supply power to the system. A DC barrel jack was used for main voltage input for the whole system. This means that an end user can use either an AC-DC adapter or a battery pack with DC jack or any DC source with a DC jack. This gives the end user the flexibility of powering the system with either a battery or any other DC source. To meet the supply voltage input requirement of WMCU as well as air temperature and humidity sensor which is 3.3 V, the LM317, an adjustable positive-voltage regulator, was used to convert the input voltage to 3.3 V. A voltage divider circuit along with the LM317 was used to convert the input voltage to 3.3 V. The LM7805, a fixed-voltage integrated-circuit voltage regulator was used to convert the main input to 5 V to satisfy the input voltage requirement of the air quality sensor. Input-bypass and output capacitor were used to ensure stable performance of the voltage regulators. The power management system design ensures that stable and appropriate power is supplied to each subsystem.

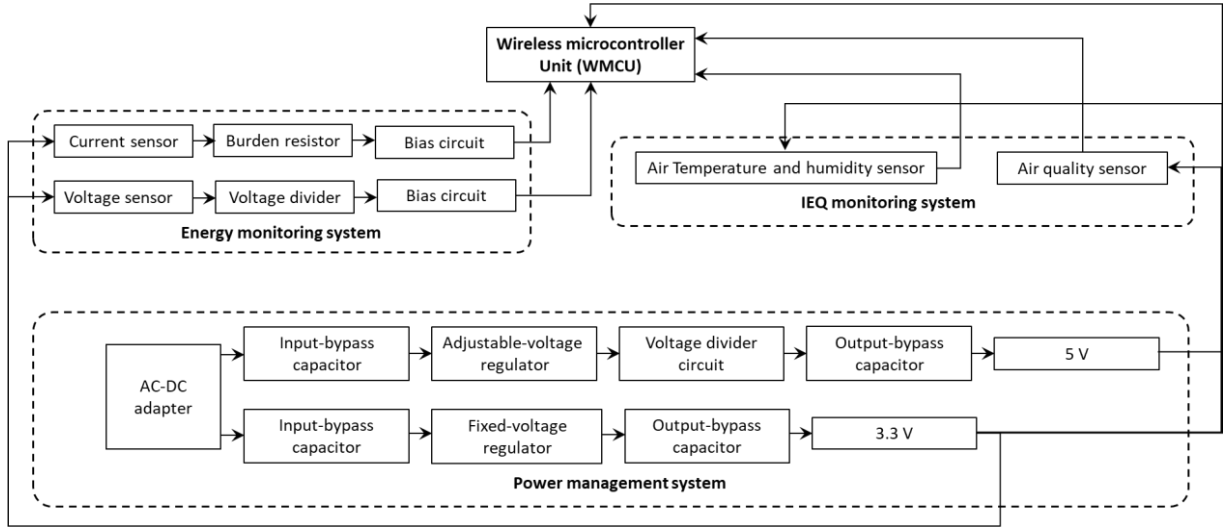

Fig. 2: Hardware design

## 3.2 Software design

### 3.2.1 Embedded software design

The embedded software design handles the data communication between all the subsystems. The WMCU was programmed according to the application requirements for proper functioning of the whole system. The design consists of the following major routines: initialization, network configuration, data acquisition, and hypertext transfer protocol (HTTP) post request routine. The initialization routine is performed only once when the system is powered on or after reset. The remaining routines run continuously as long as the system remains powered on. Library importation, pin initialization, and declaration of variables are performed in the initialization routine. Network configuration routine uses the Wi-Fi and domain (webpage account) credentials to establish network connection with Wi-Fi and the domain name used in this study. The routine continuously checks for Wi-Fi connection to ensure that it is always connected and also includes a function that tries to reconnect to the Wi-Fi in case of connection failure. The domain application program interface (API) is used to connect to the domain name where the HTTP post request will be sent. The data acquisition routine reads and computes the data from different sensors in the energy and IEQ monitoring subsystems. The output from the sensors are processed through the analog digital converter (ADC) of the WMCU, and the processed data are computed and sent to the HTTP post request routine. The HTTP post request routine receives the computed data from the data acquisition routine and transmit it to the domain database where it will be stored for further analysis. The data acquisition routine performs the following function to acquire the relevant data from each sensor. For measurement of air quality, the $CO_2$ concentration in the air measured by the air quality sensor is read by the analog pin of the WMCU and analyzed to ascertain the air quality. Similarly, through the Serial Data (SDA) and Serial Clock (SCL) of the Inter-Integrated Circuit (IIC) communication protocol in WMCU, the air temperature and humidity data measured by the air temperature and humidity sensor is read. The current and voltage signals are read by analog pins of the WMCU and the acquired data are used to compute the effective current and voltage, and consequently, active power. The mathematical equations for the computation of the aforementioned parameters are given in Eqs. (2) to (4). The effective current and voltage are defined by the square root of the mean value and are calculated according to Eqs. (2) and (3), respectively. The active power is calculated from the current and voltage in each sample according to Eq. (4). These computed measurements are then sent to the HTTP post request routine for transmission to the database of the domain name used in this work for storage and further analysis.

$$I_{RMS} = \sqrt{\frac{I_{j1}^2 + I_{S2}^2 + I_{j3}^3 + \cdots + I_{jj}^2}{N_j}} \quad (2)$$

$$V_{RMS} = \sqrt{\frac{V_{j1}^2 + V_{j2}^2 + V_{j3}^3 + \cdots + V_{jj}^2}{N_j}} \quad (3)$$

Where $I_{RMS}$ and $V_{RMS}$ are effective current and voltage, respectively, over one period. $I_{ji}$ and $V_{ji}$ are the $i^{th}$ current and voltage sample, respectively, for $i = 1, 2, 3, \ldots, s$. $N_j$ is the number of samples over one period.

$$P = \frac{1}{N} \sum_{j=1}^{N} i(j) \cdot v(j) \tag{4}$$

Where $P$ is the active power. $N$ is the number of samples. $i(j)$ is sampled instance of $i(t)$. $v(j)$ is sampled instance of $v(t)$.

### 3.2.2 Web-based information provision system

As shown in Figure 3, the information provision system employs web-based approach for effective and intuitive display of energy and IEQ data, and provision of helpful tips based on the energy consumption pattern to improve end-user's awareness on energy conservation and healthy living. The most crucial issue for the information provision system in this study is the end-user awareness and the immediacy of the information aiming at helping users cultivate energy saving and healthy living habit. A detailed, frequent, and intuitive feedback is therefore of a vital importance to help end-users cultivate the desired habit. Firstly, real-time energy and IEQ data will be plotted on the user's personal dashboard which a user can login and access the information. Secondly, when a sudden unusual pattern is observed in the monitored data, the user is notified immediately via email for timely action to be taken along with helpful tips for energy savings and healthy living.

To achieve the aforementioned function, MySQL database was created in the cPanel of the domain name used in this work. MySQL table to hold the energy and IEQ data was then created in the database. A Hypertext Preprocessor (PHP) script was created to receive the incoming requests from the HTTP post request routine and inset the data into the MySQL database. The PHP script also analyzes the data to check for any unusual pattern in the data and immediately notifies the user via email when such pattern is detected. Lastly, another PHP script was created to plot the database content on the webpage. The script also sends the daily summary information via email to the user. A user can then use their login credentials to access the information on the webpage.

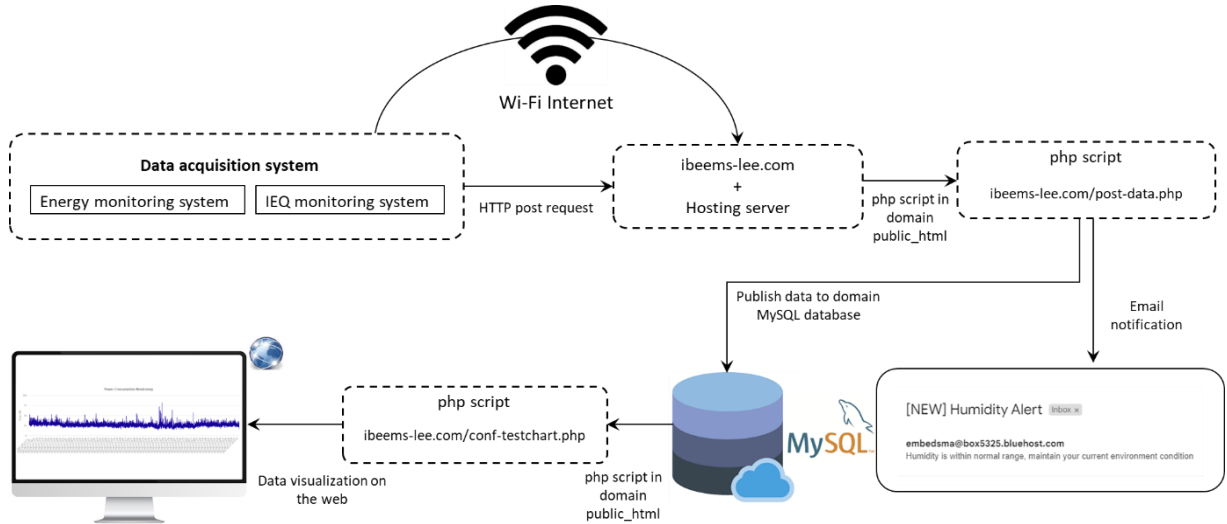

Fig. 3: Web-based information provision system

### 3.3 Working principle of the proposed system

As described in section 2.1, the proposed system is comprised of energy monitoring system, IEQ monitoring system, WMCU, power management system, and information provision system. The realized integrated hardware design is shown in Figure 4. After the system is connected to the power source and the embedded software program described in section 3.2.1 is loaded to the system, the WMCU connects to Wi-Fi and starts collecting data from the energy and IEQ monitoring systems. At the same time, the WMCU connects to the domain and start sending HTTP requests. Similarly, the PHP script created in the domain for receiving HTTP requests starts receiving the data from the WMCU and stores them in the MySQL database. And finally, the database content is visualized on the webpage by a PHP script. For energy and IEQ data acquisition, they system was configured for a one-second interval, which means that measurement was taken and transmitted to the database at every second. Furthermore, each recorded data series was labelled with a time stamp and the real-time data can be observed on the webpage. The system analyzes the energy and IEQ data and sends helpful tips to the end user to help conserve energy and maintain healthy lifestyle. If the energy consumption, air temperature, humidity, or air quality values are higher than the normal/acceptable value, the user will be notified along with suggestions on how to improve. Similarly, if the data are within the acceptable range, the user will be notified to maintain the energy consumption and environmental condition. Furthermore, if the VRMS is higher or lower than the normal value during any time of the day which indicates a voltage swell or sag, the user will immediately be alerted. Through this process, the end user is therefore well informed about energy consumption and IEQ pattern.

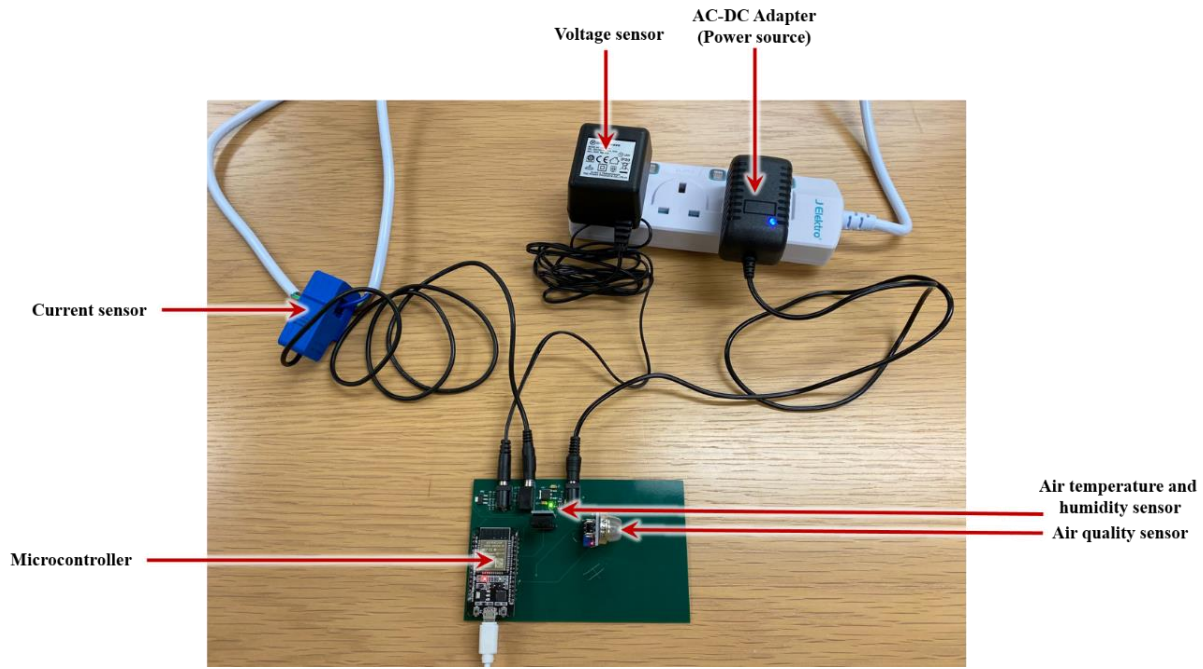

Fig. 4: Prototype of the designed energy and IEQ monitoring system

## 4. VALIDATION OF THE PROPOSED SYSTEM

### 4.1 Measurement accuracy of energy and IEQ monitoring systems

To validate the effectiveness of the proposed system, the measurements from the system were compared with commercial devices and the percentage error was computed as shown in Tables 1 and 2. The measurements from a commercial energy meter that measures current, voltage, and power was used as a reference to compare with the measurements of the energy monitoring system developed in this study. Similarly, a commercial air quality meter (i.e., RS Pro air quality meter) that measures air temperature, humidity, and air quality ($CO_2$ concentration) was used to compare the measurements from the IEQ monitoring system developed in this study. Measurements from four electrical appliances were taken by the commercial energy meter and the proposed energy monitoring system as shown in Table 1. As seen from the result, the voltage measurement was stable for all the appliance as at the time the measurements were taken for both devices and the error was 0.45%. For current measurements, the largest error was 0.56% from electric iron while the laptop and portable fan showed the same measurement values for both meters (% Error = 0). The highest error for power measurement was 0.6% and the lowest was 0.05% from electric iron and laptop measurements, respectively. Similarly, air temperature, humidity, and $CO_2$ concentration measurements were taken by the commercial air quality meter and the proposed IEQ monitoring system and results are shown in Table 2. From the result, the highest error was 0.75% and the lowest was 0.18% from air temperature and $CO_2$ concentration measurements, respectively. The validation results of the energy and IEQ monitoring systems show the competitiveness of the proposed system.

Table 1: Validation results of energy monitoring system

| Appliance | Power | | | Voltage | | | Current | | |
|---|---|---|---|---|---|---|---|---|---|
| | Reference (W) | Measured (W) | % Error | Reference (V) | Measured (V) | % Error | Reference (A) | Measured (A) | % Error |
| Kettle | 1653.9 | 1648.69 | 0.32 | 223 | 222 | 0.45 | 7.40 | 7.38 | 0.27 |
| Laptop | 38.00 | 37.98 | 0.05 | 223 | 222 | 0.45 | 0.17 | 0.17 | 0 |
| Electric iron | 1202.43 | 1195.19 | 0.60 | 223 | 222 | 0.45 | 5.38 | 5.35 | 0.56 |
| Portable fan | 24.59 | 24.57 | 0.08 | 223 | 222 | 0.45 | 0.11 | 0.11 | 0 |

Table 2: Validation results of IEQ monitoring system

| IEQ | Reference | Measured | % Error |
|---|---|---|---|
| Air temperature (°C) | 26.7 | 26.9 | 0.75 |
| Humidity (%) | 56.3 | 56.1 | 0.36 |
| $CO_2$ concentration (ppm) | 569 | 568 | 0.18 |

## 4.2 Real-time energy and IEQ data monitoring

The web-based real-time data monitoring functionality was validated through a five-hour long experiment where the proposed device was used to monitor the environment and energy consumption of a participant. The participant used a portable fan to keep the environment thermally comfortable and the current consumed was measured using the current sensor. Similarly, voltage was measured with the voltage sensor and power was computed from the current and voltage data. The air temperature, humidity, and air quality ($CO_2$ concentration) were also measured using the environmental sensors. The data were collected, computed, published to the database, and finally displayed on the web according to the method described in section 3.2.1 and the result is shown in Figure 5. As shown in Figure 5 (D), (E), and (F), the measurements fluctuate between 24.2°C and 24.6°C, 55.25% and 56%, and 564 ppm and 568 ppm for air temperature, humidity, and $CO_2$ concentration, respectively. Throughout the experiment, the measurements are consistent within the above-mentioned range as seen in Figure 5 (D), (E), and (F). High and irregular fluctuations can be seen from the measurements of current, voltage, and consequently power as shown in Figure 5 (A), (B), and (C). This could be attributed to the power supply noise during the time of the experiment. From the result, an end user will be able to intuitively observe the energy consumption and IEQ pattern of the monitored environment.

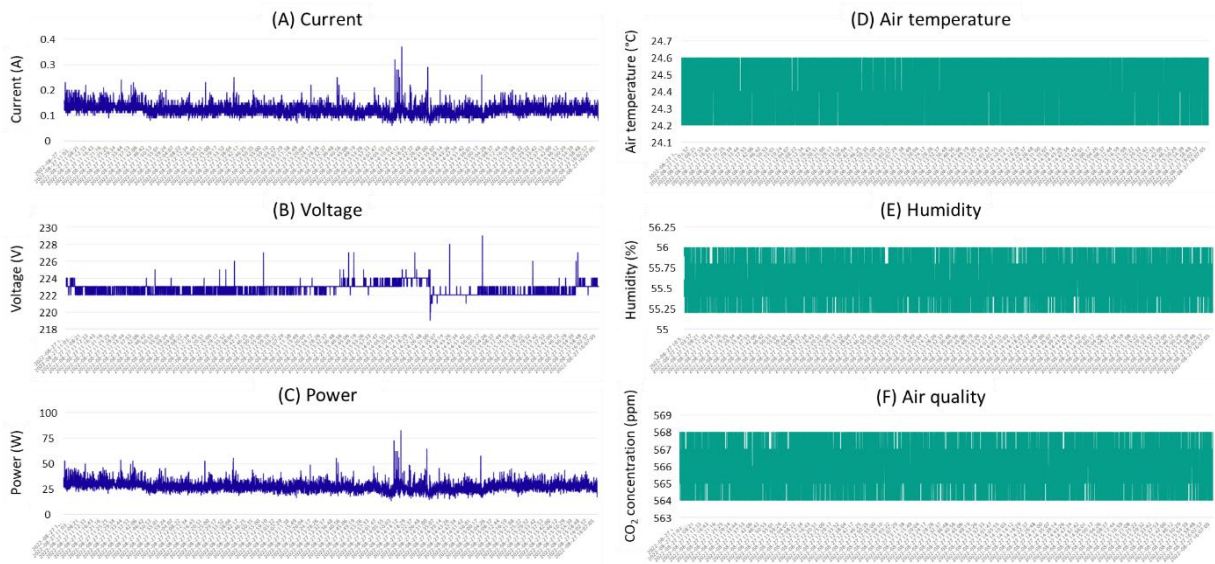

Fig. 5: Real-time energy and IEQ data monitoring

### 4.3 Notification email by information provision system

The information provision function of the proposed system was validated during the experiment described in section 4.2. As the data were collected and transferred to the database, they were analyzed to check for patterns according to the method described in section 3.2.2. When a certain pattern is observed, the user is notified immediately via email with information regarding the observed pattern. For this experiment in particular, the system was to check for: (i) whether the indoor environmental condition is within the normal/acceptable range or not; (ii) voltage swell in the power system which is a sudden increase in voltage; and (iii) voltage sag which is a sudden decrease in voltage supply. In this experiment, when the voltage reading is above 225 V, it is considered as voltage swell and when it is below 219 V, it is considered as voltage sag. Voltage swell and sag which are related to power quality issues were analyzed as they can cause an instantaneous power consumption increase/decrease and could lead to destruction of electrical appliances. Since the air temperature, humidity, and air quality were within the normal range as at the time of the experiment as seen in section 4.2, the user was notified and advised to maintain the environmental condition as shown in Figure 6 (A), (B), and (C). As seen in section 4.2, it can be observed that voltage swell and sag had occurred at certain times during the experiment. The user was therefore notified about this power quality issue and advised to check the power system as shown in Figure 6 (D) and (F). Thus, the user is updated about the monitored environmental condition and potential issues with the power system and consequently power consumption.

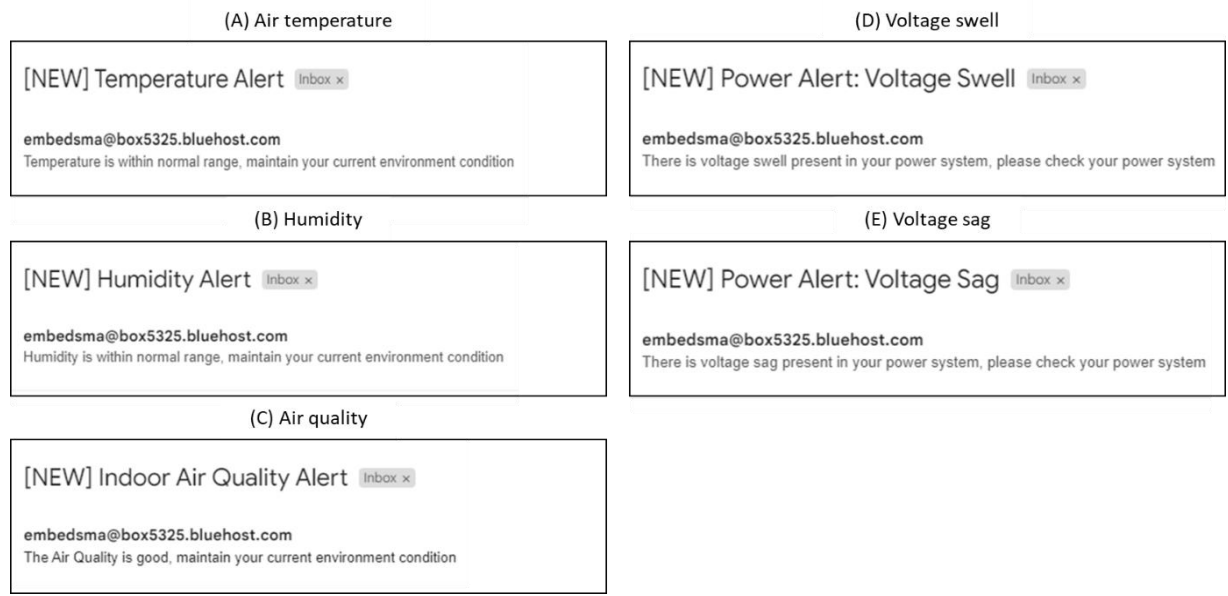

Fig. 6: Notification email by information provision system

### 5. CONCLUSION

This study designed and proposed an integrated solution for energy and IEQ monitoring towards energy conservation and healthy living. The study aims to realize an equilibrium between energy saving behavior and healthy living since one is often sacrificed to gain the other. The system is a hybrid device that comprise of subsystems each performs different functions for the realization of the overall aim of the system. Through the systematic design approach, each subsystem though designed differently were integrated to form a compact and user-friendly system. The user friendliness which was at the core of the design aims to allow users with little technical knowledge to be able to operate the system conveniently. The hardware and software design were both realized which will give an end user a one in all solution for energy and IEQ monitoring. The system was validated through different experiments and the results demonstrates the competitiveness and usefulness of the proposed system in energy and IEQ monitoring. The proposed system has potential for development of intuitive application to encourage occupant's energy consumption and healthy behavioral change. Therefore, our future work will utilize this designed system to analyze the correlation between energy consumption and IEQ data, identify energy saving potentials while maintaining a satisfactory IEQ, and design an information feedback system to help end users and building management to facilitate the adoption of energy conservation and healthy living lifestyle, while this study mainly focused on the realization of functions required for these applications. This study ultimately intends to advocate energy conservation habit that does not comprise the health of the occupant.